# Fast Neural Network Approach for Direct Covariant Forces Prediction in Complex Multi-Element Extended Systems

**Jonathan P. Mailoa,**[1*] **Mordechai Kornbluth,**[1] **Simon L. Batzner,**[2,3] **Georgy Samsonidze,**[1] **Stephen T. Lam,**[1,2] **Chris Ablitt,**[1,4] **Nicola Molinari,**[1,3] **and Boris Kozinsky**[3,1*]

1) Bosch Research and Technology Center, Cambridge, MA 02139, USA
2) Massachusetts Institute of Technology, Cambridge, MA 02139, USA
3) Harvard School of Engineering and Applied Sciences, Cambridge, MA 02138, USA
4) Imperial College London, London SW7 2AZ, UK

* corresponding author: jpmailoa@alum.mit.edu, bkoz@seas.harvard.edu

# Fast Neural Network Approach for Direct Covariant Forces Prediction in Complex Multi-Element Extended Systems

**Abstract**

Neural network force field (NNFF) is a method for performing regression on atomic structure – force relationships, bypassing expensive quantum mechanics calculation which prevents the execution of long ab-initio quality molecular dynamics simulations. However, most NNFF methods for complex multi-element atomic systems indirectly predict atomic force vectors by exploiting just atomic structure rotation-invariant features and the network–feature spatial derivatives which are computationally expensive. We develop a staggered NNFF architecture exploiting both rotation-invariant and covariant features separately to directly predict atomic force vectors without using spatial derivatives, thereby reducing expensive structural feature calculation by ~180–480×. This acceleration enables us to develop NNFF which directly predicts atomic forces in complex ternary and quaternary–element extended systems comprised of long polymer chains, amorphous oxide, and surface chemical reactions. The staggered rotation-invariant-covariant architecture described here can also directly predict complex covariant vector outputs from local physical structures in domains beyond computational material science.

## Introduction

Ab-initio molecular dynamics (AIMD) is an atomistic simulation method widely used to study the movement of atoms in a physical system, where the forces experienced by each atom in the system is calculated using a quantum mechanics method such as density functional theory (DFT). Quantum interatomic force calculations are produced by solving a many-body system including electrons (e.g. Schrödinger, Kohn-Sham equations). The computational cost of these methods make AIMD computationally challenging for realistic physical phenomena that can be explored only when the simulated system is sufficiently large (many atoms) and/or has run for a long time (many time steps). However, AIMD is still widely used (despite the limitations in size and time scale) because it requires no prior assumption on the potential energy surface, and it can be used to accurately simulate interesting phenomena such as chemical reactions, phase changes, ionic transport, surface interactions, etc in a wide variety of material systems.

Classical molecular dynamics (MD) methods based on fast force calculations using pre-fitted empirical functions are $10^5$–$10^6$ × faster than AIMD, but the limitations of simple empirical functions often mean that they cannot be used to study complex atomic interactions, e.g. chemical reactions. Machine learning methods have increasingly been used to perform atomistic computations of energies and atomic forces with greater accuracy than empirical functions.[1–11] Some of these latest methods have shown high force prediction accuracy (error within 1 kcal/mol·Å = 0.043 eV/Å) for single-molecule systems in vacuum,[12,13] but may be unsuitable for extended solid-state atomistic systems containing large number of atoms. Other kernel-based methods such as the Gaussian process regression have been used for developing force fields for single-element nanocluster in vacuum (force error within 0.20 eV/Å for many-body kernels).[14] The neural network force field (NNFF) utilizes flexible neural network (NN) functions at fixed computation cost (independent of training sample size), which in turn enables indirect atomic

structure – force regression involving many-body interactions, and has been used for complex multi-element extended systems.[15–17] Due to the vast number of possible atomic configurations in 3-dimensional space, NNFF models were difficult to train.[18] The development of 'atomic fingerprints', by Behler and Parrinello (B–P) and others,[15,19] has enabled a step-change improvement in the accuracy of NNFF. Similar to convolutional and graph neural network (CNN and GNN),[20–22] the atomic fingerprints rely on features obtained from localized proximity in space. Unlike CNN and GNN which utilize fixed-size grid-based and graph-based feature space respectively, Behler-Parrinello style atomic fingerprints use a fixed-size symmetry-function-based feature space utilizing all atoms located at $\vec{r}_j$ within a specified cutoff radius $R_c$ from the target central atom $i$ ($\{\vec{R}_i\} = \{\vec{r}_j$ for $|\vec{r}_{ij}| < R_c, j \neq i\}$). These symmetry functions $\vec{G}_i$ work well for describing $\{\vec{R}_i\}$ because they exhibit translational, rotational, and same-element-permutation invariance, as well as smoothly decaying contributions from atoms farther away from the central atom.[15] Equations (1-3) are examples of standard B–P symmetry functions, where $f_c$ is cutoff function, $r$ is distance, $G$ is symmetry function, $\alpha, \beta$ are atomic elements, $\vec{r}_{ij}$ is relative position of atom $j$ from atom $i$, $\cos\theta_{ijk} = \vec{r}_{ij} \cdot \vec{r}_{ik}/|\vec{r}_{ij}||\vec{r}_{ik}|$, and $\eta, R_s, \zeta, \lambda$ are the function parameters. The illustration of the B–P atomic fingerprint scheme is shown in **Figure 1a**.

$$f_c(r) = \begin{cases} 0.5 \cdot [\cos(\pi r/R_c) + 1] & \text{for } r \leq R_c \\ 0 & \text{for } r > R_c \end{cases} \qquad (1)$$

$$G^\alpha_{2-body}(\{\vec{R}_i\}) = \sum_{j \in \alpha} e^{-\eta(|\vec{r}_{ij}|-R_s)^2} f_c(|\vec{r}_{ij}|) \qquad (2)$$

$$G^{\alpha,\beta}_{3-body}(\{\vec{R}_i\}) = 2^{1-\zeta} \sum_{\substack{j \in \alpha \\ k \in \beta \\ j \neq k}} (1 + \lambda \cos\theta_{ijk})^\zeta e^{-\eta(|\vec{r}_{ij}|^2 + |\vec{r}_{ik}|^2 + |\vec{r}_{jk}|^2)} f_c(|\vec{r}_{ij}|) f_c(|\vec{r}_{ik}|) f_c(|\vec{r}_{jk}|) \qquad (3)$$

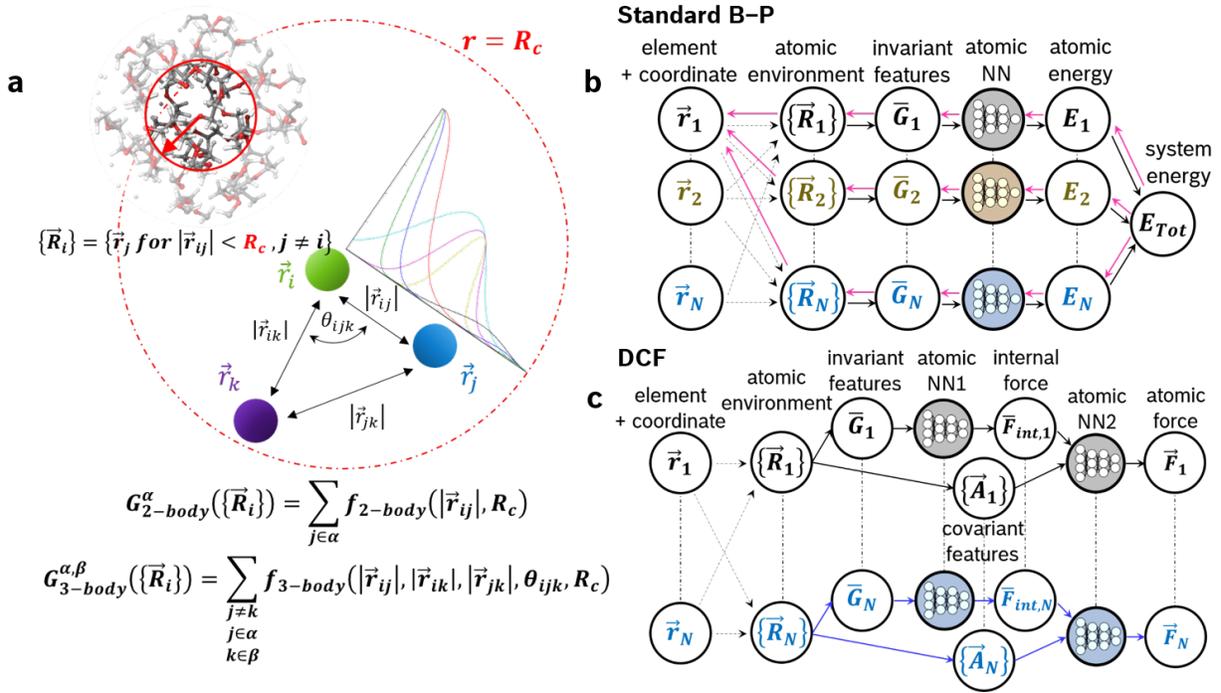

**Figure 1**. a) Behler – Parrinello local atomic environment rotation-invariant feature extraction scheme. b) Indirect force prediction in standard Behler–Parrinello (B–P) NNFF scheme from network and feature differentiation. Pink arrow indicates the differentiation computation pathway needed for just one of the atoms in the system. c) Direct covariant forces (DCF) prediction scheme utilizing rotation-invariant and covariant features which bypass the need for computationally intensive feature differentiation.

Nevertheless, NNFFs based exclusively on rotation-invariant atomic fingerprints (standard NNFF) are still computationally expensive for atomic force calculations. Because the input atomic fingerprints are rotationally-invariant (contains no directional information), these features cannot be used to perform regression on Cartesian force vectors outputted by a quantum mechanical simulation (contains directional information). These fingerprints are instead used to perform neural network regression on rotationally-invariant outputs such as energy, followed by differentiation of the energy with respect to the Cartesian space to obtain the Cartesian force vectors. Unfortunately, in practice this requires the derivative of the energy neural network with respect to all of the atomic fingerprints, followed by differentiation of each of the atomic fingerprints with respect to all of the atomic positions within the range of the fingerprint function's cutoff radius $R_c$ as shown in **Figure 1b**. The computational cost of atomic fingerprints is high, especially if including 3-body terms, and it constitutes the computation bottleneck of each NNFF MD time step. Having to perform differentiation of those fingerprints multiplies the computation cost by approximately $3m$, with $m$ being the average number of atoms within each fingerprint function's $R_c$. For fingerprint functions with $R_c$ = 6.5 Å within AIMD simulations in this work, $m$ is in the range of 58–157, which means the fingerprint derivative operation slows down NNFF training and test evaluation time by ~180–480× per iteration, depending on the environment $\{\vec{R}_i\}$ under evaluation. Furthermore, because the NNFF output (system energy) is a combined property of all the atoms in the system, atomic force training and prediction for different atoms are coupled. If the force prediction on a rare atomic event of interest (a few atoms in a chemical reaction, for example) is inaccurate, the training set can only be improved by adding entire simulation frames which contains many additional $\{\vec{R}_i\}$ of little interest. It is worth noting that alternatives to the B–P approach have been formulated that bypass the explicit construction of the symmetry functions and rely instead on deeper network architectures;[23,24] these approaches also predict total energies of molecules and also require differentiation of the entire network to calculate the forces.

In this study, we present a neural network based approach for direct covariant forces, or DCF, that enables prediction of Cartesian force vectors in extended solid-state systems from just multi-element local atomic environments while maintaining the benefit of rotation-invariant atomic fingerprint features and bypassing the requirement to perform expensive Cartesian space derivatives of these atomic fingerprint features. By its nature, this means the training and prediction of atomic forces for atoms in our DCF algorithm will be significantly faster than in the standard NNFF algorithm (~180–480x speedup expected) and requires only local atomic environments. We will further discuss the benefits of being able to decouple total system energy from the force prediction algorithm at the end of this manuscript. The acceleration in DCF enables us to develop NNFF which directly predicts atomic forces in complex ternary and quaternary–element extended systems comprised of long polymer electrolyte chains, amorphous oxide electrolyte, and surface chemical corrosion reactions. Beyond the domains of computational materials science, the approach described here is relevant for machine learning in other physical systems where the rotational symmetry of physical laws can be exploited for vector output prediction, e.g. fluid dynamics[25,26] and mass/heat transport.[27]

*Notation*: We use an arrow vector $\vec{x}$ to refer to a quantity that is directional in Cartesian space, e.g. position, force, or a reference axis. A straight-line vector $\bar{x}$ refers to an array of scalars, e.g. a list of coefficients or a set of fingerprints for a given atom. The two together, i.e. an array of directional vectors, is represented as $\{\vec{x}\}$, e.g. a set of positions or a list of forces.

## Direct Covariant Forces Prediction Algorithm

The general architecture of the NN capable of performing direct force vector prediction from a local atomic environment is shown in **Figure 1c**, which is enabled by the inclusion of rotation-covariant atomic fingerprints $\{\vec{A}_i\}$ in addition to the rotation-invariant fingerprints $\bar{G}_i$. The key component of this approach is the usage of rotation-invariant features to predict a smaller number of rotation-invariant intermediate hidden states, followed by later inclusion of rotation-covariant features to predict the final rotation-covariant output vector. Physically, $\{\vec{A}_i\}$ describes internal orientation reference axes for atom $i$'s $\{\vec{R}_i\}$. As shown in equation (4), $\{\vec{A}_i\}$ (unlike $\bar{G}_i$) encode directionality information. Just like $\bar{G}_i$, $\{\vec{A}_i\}$ are smoothly decaying functions of $\{\vec{R}_i\}$ exhibiting input translation and same-element permutation invariance (except for when there is only one atom $j$ of element $\alpha$ entering/exiting $\{\vec{R}_i\}$). However, $\{\vec{A}_i\}$ features are rotation-covariant in nature, defined as the commutation of the rotation transformation $T$ with the dependence of $\{\vec{A}_i\}$ on atomic positions $\{\vec{R}_i\}$: $\{\vec{A}_i\}(T\{\vec{R}_i\}) = T\{\vec{A}_i\}(\{\vec{R}_i\})$. One example is given in equation (4-5).

$$\vec{V}^{\alpha}_{2-body}(\{\vec{R}_i\}) = \sum_{j \in \alpha} \vec{r}_{ij} e^{-\eta(|\vec{r}_{ij}|-R_s)^2} f_c(|\vec{r}_{ij}|) \tag{4}$$

$$\vec{A}^{\alpha}_{2-body}(\{\vec{R}_i\}) = \begin{cases} 0 & \text{for } \vec{V}^{\alpha}_{2-body} = 0 \\ \vec{V}^{\alpha}_{2-body}/|\vec{V}^{\alpha}_{2-body}| & \text{otherwise} \end{cases} \tag{5}$$

Once we have these rotation-invariant and rotation-covariant features defined, we start building our NNFF in a decoupled manner to ease the learning process. First, we build a rotation-invariant network NN1 for each atom element consisting of exclusively $\bar{G}_i$ as its input (**Figure 2**). Because $\bar{G}_i$ is rotation-invariant, NN1 can only be used to perform regression on rotation-invariant outputs. Both the atomic forces on atom $i$ ($\vec{F}_i$) and internal axes $\{\vec{A}_i\}$ are rotation-covariant. However, the projection of $\vec{F}_i$ onto $\{\vec{A}_i\}$ is rotation-invariant.[28] Hence we can define a rotation-invariant value 'internal force' $\bar{F}_{int,i} = \{\vec{A}_i\} \cdot \vec{F}_i$, which we can use as the intermediate output of NN1, and perform $\bar{G}_i \rightarrow \bar{F}_{int,i}$ regression training. As it turns out, $\bar{G}_i$ can contain sufficient physical information to describe the quantum-mechanical output $\bar{F}_{int,i}$ which is directly related to the rotation-covariant value $\vec{F}_i$ we are interested in (see Supplementary **Figure S1**).

The subsequent step is in contrast to the pseudoinverse approach taken in previous Gaussian process work,[28] which tends to diverge when $\{\vec{A}_i\}$ does not span the full orthogonal 3-dimensional space (see Supplementary Information). In order to transform rotation-invariant value $\bar{F}_{int,i}$ back into the Cartesian space vector $\vec{F}_i$, we build a second network NN2 with $\bar{F}_{int,i}$ and $\{\vec{A}_i\}$ as the input and $\vec{F}_i$ as the output (**Figure 2**). Unfortunately none of the widely-used hidden nodes available in the Python TensorFlow package have 3D-rotation-covariance built in. Because of that, the practical solution we have chosen to enforce rotation-covariance on NN2 is to perform data-augmentation. In addition to having $(\bar{F}_{int,i}, \{\vec{A}_i\}) \rightarrow \vec{F}_i$ in the training set, we also include $(\bar{F}_{int,i}, T\{\vec{A}_i\}) \rightarrow T\vec{F}_i$ pairs, where $T$ is a random 3D-rotation matrix. In principle, one can augment the data with many different $T$ matrices to improve data augmentation. In practice, we have chosen to incorporate as many unique $(\bar{F}_{int,i}, \{\vec{A}_i\}) \rightarrow \vec{F}_i$ pairs as possible and only augment each data point with an appropriate number of $(\bar{F}_{int,i}, T\{\vec{A}_i\}) \rightarrow T\vec{F}_i$ replicated pairs depending on the data size of the unique pairs (10 augmentation for system A, 1 augmentation for system B and C). Unlike NN1 which learns to interpolate quantum mechanical forces using high-dimensional fingerprints $\bar{G}_i$, the role of NN2 is simply to perform geometric inverse projection using much lower dimensional inputs $(\bar{F}_{int,i}, \{\vec{A}_i\})$. Similarly, we show that $(\bar{F}_{int,i}, \{\vec{A}_i\})$ contains sufficient information to perform regression on $\vec{F}_i$ (see Supplementary **Figure S2**). We note that this is not a true covariant operation, and a better approach should be implemented in the future to enforce true covariance in the model upon inclusion of covariant features. We note that recent approaches were proposed in this direction to directly predict truly covariant force vectors, but they require implementation of new custom-built 3D covariant neural network nodes and as a result are computationally and practically more complex, and have not been demonstrated in extended systems.[29,30]

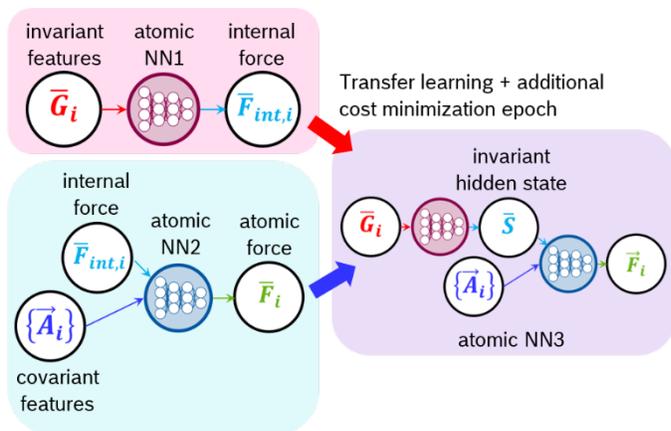

**Figure 2**. DCF network NN3 composed of rotation-invariant (NN1) and rotation-covariant (NN2) components of the network were trained independently using rotation-invariant scalars $\bar{F}_{int,i}$ as the intermediary nodes. NN2 is biased toward becoming like a rotation-covariant function using data augmentation.

Finally, we build the final network combining the architecture of NN1 and NN2 (**Figure 2**). This full network NN3 has the exact sub-geometries of NN1 and NN2, where the nodes representing $\bar{F}_{int,i}$ now appearing as hidden states. After performing transfer learning to move the trained weights and biases from NN1 and NN2 into NN3, we perform additional training cycles on NN3 using the $(\bar{G}_i, \{\vec{A}_i\}) \to \vec{F}_i$ dataset. Effectively, what we have done is to let NN1 learn quantum mechanics, NN2 learn inverse geometry projection, and NN3 minimizes the regression error from the NN1-NN2 combined system. In the beginning of the training, NN3 is initiated with $\bar{F}_{int,i}$ as target values for the rotation-invariant hidden states but is eventually allowed to find different hidden states which locally minimize the error. See supplementary **Table S1–S3** for the network architectures and the set of hyperparameters we have chosen for $\bar{G}_i$, $\{\vec{A}_i\}$, and $R_c$ in this work. More importantly, because there is no need to take the derivative of the entire model to compute forces, expensive fingerprint calculation only needs to be performed once for each $\{\vec{R}_i\}$ during both training and MD runtime instead of ~$3m$ times needed for training of the original B–P algorithm.

## Evaluation of DCF on Complex Atomistic Systems

We now investigate the performance of DCF on the test set. We use three complex atomistic systems to evaluate the performance of DCF. The first system (A, **Figure 3a**) is a polyethylene oxide (PEO) polymer consisting of 10 chains of 100 monomers each. PEO is an important material because it is the polymer electrolyte most widely used for solid-state Lithium-polymer battery applications.[31,32] Long-time polymer dynamics are difficult to study using ab-initio MD because of the large number of atoms necessary to represent the amorphous structure of the polymer in a periodic simulation box. The polymer geometry snapshots were obtained using classical MD, but the $\vec{F}_i$ outputs were then computed using quantum mechanics on the extracted mini-snapshots. This polymer system contains 7090 atoms per MD frame (2020 C, 4060 H, and 1010 O atoms) and was run at temperature $T$ = 353 K (typical polymer electrolyte operating temperature). The second system (B, **Figure 3b**) is amorphous lithium phosphate ($Li_4P_2O_7$) oxide, a promising Li-ion conducting solid electrolyte. The structure consists of 208 atoms (64 Li atoms, 112 O atoms, and 32 P atoms) melted at $T$ = 3000 K. This ab-initio MD melt simulation is interesting because it represents the rapid thermal annealing process which can be used to turn crystalline $Li_4P_2O_7$ (poor Li conductor) into amorphous $Li_4P_2O_7$ (better Li conductor),[16] where different phases of the oxide are present during the simulation. The third system (C, **Figure 3c**) is the surface of aluminum oxide ($Al_2O_3$) in the presence of high concentration of hydrofluoric acid (HF) gas. This structure consists of 228 atoms (64 Al, 34 F, 34 H, and 96 O atoms) and was run at $T$ = 1000 K. This system is interesting because during the simulation, the $Al_2O_3$ surface gets corroded by the chemical reaction with HF molecules. Understanding such corrosion reactions is important in many technological applications, such as proton exchange membrane (PEM) fuel cell bipolar plate corrosion which significantly reduces fuel cell lifetime.[33] See Methods & Supplementary Information for details on quantum mechanical simulation & data extraction.

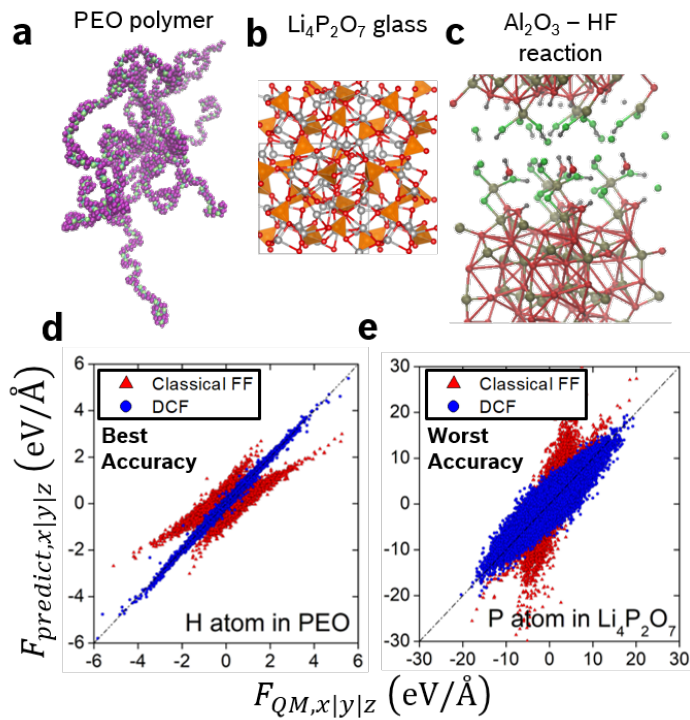

**Figure 3**. Structures being used for DCF training: a) amorphous polyethylene oxide (PEO) polymer electrolyte, b) amorphous $Li_4P_2O_7$ oxide electrolyte, c) $Al_2O_3$ surface reaction with HF molecules. The DCF force prediction correlation plots (blue dots) are shown for: d) H atoms in PEO which are the easiest atoms to perform regression on, and e) P atoms in $Li_4P_2O_7$ which are the hardest atoms to perform regression on. Classical FF force predictions (red triangles) are shown for comparison.

One of the first distinct observations we learned from the DCF algorithm is that different atoms have different levels of learning difficulty. Forces for atoms in relatively simpler chemical environments, such as single-bonded H or F atoms are relatively easy to learn using smaller number of atomic environments. This can immediately be seen in **Figure 3d**, where the blue data points corresponding to DCF $\vec{F}_i$ regression on the test set showing very good force accuracy for H atoms in system A. On the other hand, atoms in more complex chemical environments, such as P atoms in tetragonal $PO_4$ sites in system

B, are significantly more difficult to train the NNFF on. As shown by the blue DCF regression data points in **Figure 3e**, while there is a clear correlation trend of predicted $\vec{F}_i$ with input quantum mechanical forces, it is clear that such system can be improved further to reach the level of accuracy seen in **Figure 3d**.

We evaluate the force prediction accuracy of DCF for system A, B and C in **Table 1**. For reference, we include the prediction accuracy of classical force field (FF) when available. For system A reference, we use the commercial OPLS 2005 force field[34] implemented in the Schrodinger software package which is primarily used for simulating organic materials. For system B reference, we use a force field from the literature developed for oxide systems.[35] There is no classical force field available for simulating $Al_2O_3$ – HF chemical reaction, so the corresponding blocks in **Table 1** has been left blank. To the best of the authors' knowledge, the only alternative form of classical force field capable of simulating chemical reaction is the ReaxFF force field.[36] Such force fields are very complicated to hand-craft and parametrize, and none has been developed for surface chemical reactions including the $Al_2O_3$–HF reaction. From **Table 1** it can be seen that the DCF algorithm has lower mean-absolute-error (MAE) force error prediction compared to the off-the-shelf classical FF. More comprehensive error evaluation for these three systems are shown in Supplementary Information.

For system A, the training data source comes from classical MD geometry snapshots separated by 5 ns intervals (long MD time scale) over a 200 ns long simulation, making the training data snapshots geometrically more diverse by construction. Because the QM simulations for system A are performed on individual atomic environments, we were only able to generate ~40,000 DFT calculations for each of C, H, and O atoms. On the other hand, the training data snapshots for systems B and C are separated by 2.0 fs and 0.5 fs (short MD time scale), respectively. Significantly more data points can be generated for these systems because the QM simulations are done on full-system snapshots, but the training data is geometrically less diverse because the simulated physical time is short.

| System | Atom | Sample # | | FF $\vec{F}_{MAE}$ (eV/Å) | NN $\vec{F}_{MAE}$ (eV/Å) | DFT $\langle|\vec{F}|\rangle$ (eV/Å) | $\vec{F}_{MAE}/\langle|\vec{F}|\rangle$ | |
|---|---|---|---|---|---|---|---|---|
| | | Train | Test | | | | FF | NN |
| (A) PEO 353 K | C | 36360 | 4040 | 1.222 | 0.361 | 2.365 | 52% | 15% |
| | H | 36540 | 4060 | 1.088 | 0.128 | 1.411 | 77% | 9% |
| | O | 39996 | 4444 | 0.93 | 0.431 | 1.736 | 54% | 25% |
| (B) Li$_4$P$_2$O$_7$ 3000 K | Li | 675000 | 75000 | 2.736 | 0.603 | 1.629 | 168% | 37% |
| | O | 675000 | 75000 | 5.521 | 0.864 | 3.446 | 160% | 25% |
| | P | 675000 | 75000 | 4.919 | 2.375 | 5.868 | 84% | 40% |
| (C) Al$_2$O$_3$ – HF 1000 K | Al | 450000 | 50000 | N/A | 0.617 | 1.846 | N/A | 33% |
| | F | 450000 | 50000 | | 0.333 | 1.525 | | 22% |
| | H | 450000 | 50000 | | 0.249 | 1.116 | | 22% |
| | O | 450000 | 50000 | | 0.538 | 1.551 | | 35% |

**Table 1**. Evaluation of force prediction accuracy for the new DCF algorithm. We define $\vec{F}_{MAE} = \langle|\vec{F}_{NN} - \vec{F}_{DFT}|\rangle$. A more complete table of the model's accuracy is shown in the Supplementary Information.

Some evaluation of standard B–P algorithm force accuracy for multi-element solid-state systems is available in the literature, but not directly comparable to this work because the training data is generated from ab-initio MD running at different or unspecified temperatures, and sometimes the error of predicted Cartesian force components is reported instead of the full force vector error.[37,38] Hence we focus on reporting a more comparable metric such as $\vec{F}_{MAE}/\langle|\vec{F}|\rangle$ rather than focusing on $\vec{F}_{MAE}$ magnitude, to be used for future comparisons of NNFF algorithms' performance. In the current implementation of DCF, this value ranges from 9% for H atoms in system A (simple single-bond –CH environment) running at 353 K to 40% for P atoms in system B (complex PO$_4$ environment) running at

3000 K. DCF prediction correlation for all atoms in system A, B, and C are included in supplementary **Figure S3**.

We also evaluate the computational speed of a DCF NN algorithm for system B, for which a comparison with both full QM (DFT) and classical FF are available, as shown in **Figure 4**. For this system, DFT QM force calculation is ~$10^6$× slower than classical FF, while the standard NNFF algorithm written in the PROPhet plugin for LAMMPS (C++) is ~350× slower than classical FF.[39] Because of the unavailability of some fingerprint functions of standard B–P in the AMP software package, we use a simplified set of $\bar{G}$ functions for directly comparing the speed of standard B–P and DCF NNFF algorithms compared to the set used for accuracy evaluation in **Table 1** (see **Table S4**). The DCF algorithm implementation today (in Python) is ~60% faster than the standard B–P algorithm implementation (in C++). While this Python DCF implementation is still much slower than classical FF in LAMMPS, further investigation of computational cost reveals that the majority of this cost is incurred due to the slow atomic fingerprint calculation code written in Python. The speed of the NN component of DCF (optimized TensorFlow library written in C++) is comparable to the speed of classical FF. In future implementations, the fingerprint calculation code will be ported directly into an MD engine such as LAMMPS (written in C++), therefore bypassing the Python computation bottleneck. Based on the available computational cost data, we can estimate that the DCF algorithm integrated into the PROPhet plugin implementation for LAMMPS (C++) will only be ~2.7× slower than classical FF (see Methods).[37,39] We again emphasize that this speedup is enabled by the elimination of fingerprint derivative calculations in an algorithm which directly predicts force vectors.

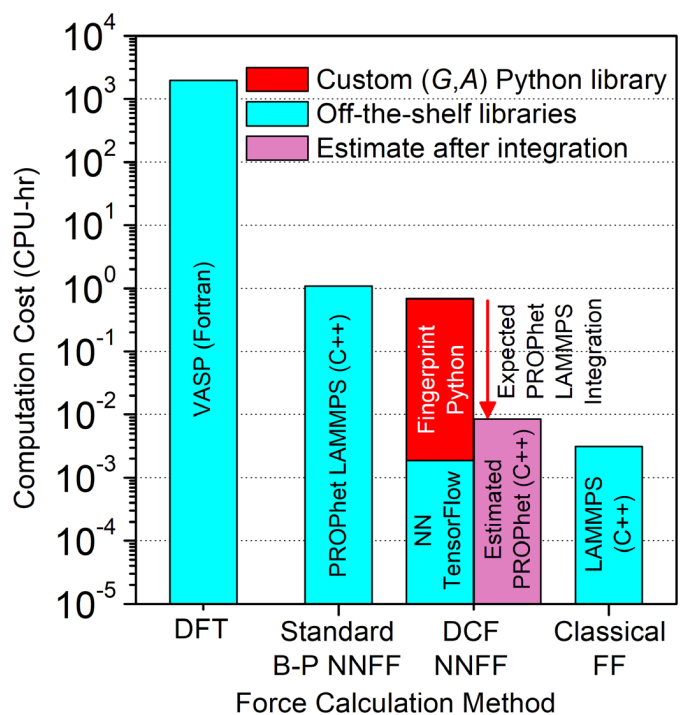

**Figure 4**. Computation cost for running molecular dynamics (MD) in system B for different methods such as DFT in Fortran, standard B–P in C++, DCF in Python (plus C++ speed estimation), and classical FF in C++. We estimate that the integration of DCF into PROPhet plugin for LAMMPS will accelerate NNFF runtime by ~127x over standard B–P scheme.

Finally, we quantify the performance of the DCF NNFF algorithm on the most difficult data set we have available, which is the reaction dataset of system C. Unlike system A and system B, which only consists of covalent bond vibration and atomic/ionic hops, system C contains chemical reactions where chemical compounds form from distinct reactants. These are rare events; there are relatively fewer atomic snapshots corresponding to the chemical reaction, making these snapshots a very small fraction of the training set data. Nevertheless, we can construct a test dataset consisting of all of the reaction snapshots from the full dataset (larger than and including a very small fraction of the training set data of system C), and perform force prediction on many reaction snapshots which have not been seen in the training set. The extraction of 'reacting atomic snapshots' from the AIMD trajectories is done using a combination of tools based on molecular graph analysis and a Hidden Markov Models (see Methods). We demonstrate this result in **Table 2**, showing the prediction of forces on atoms in system C for both standard and reaction test data set. It can be seen from **Figure 5a–b** that the performance of the NNFF trained on a typical data set performs reasonably well on the reaction data set for Al, F, and H atoms, albeit slightly worse than the performance on standard data set. In **Figure 5**, we also show some loops corresponding to a few rare reaction trajectories where the DCF fails to accurately capture forces (which is especially bad on O atoms). We hypothesize that the performance on F and H reaction snapshots are excellent because they stay in geometrically simple single-bond configurations during chemical reactions. On the other hand, the performance on O reaction snapshots is the worst because there are very few O snapshots which are involved in chemical reactions in the full dataset. These ultra-rare events constitute only ~0.065% of the entire O environment dataset, making force prediction on those chemical reaction events difficult.

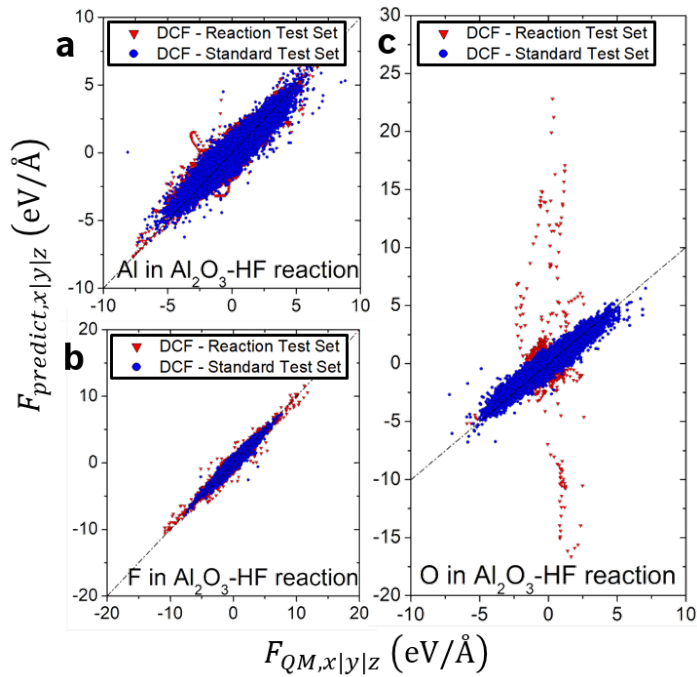

**Figure 5**. DCF force correlation plot for the 'chemical reaction test set' (red triangles) of system C, where a) Al, b) F, and c) O atoms are shown. Performance on the standard test set (clue does) is overlaid on top of the reaction test set for direct visual comparison. F and H atoms' force regressions are the best and comparable to those of the standard test set (H atoms not shown). While Al force regression on the reaction test set shows poor performance on some chemical pathways (small loops), the overall regression error is comparable to that of the standard test set. However, O atoms' regression on the chemical reaction test set are very poor compared to that of the standard test set, likely because the reaction test snapshots correspond to ultra-rare events.

| Atom | Standard Test Set | | Reaction Test Set | | Rare Event Occurrence |
|---|---|---|---|---|---|
| | Sample Count | $\vec{F}_{MAE}$ (eV/Å) | Sample Count | $\vec{F}_{MAE}$ (eV/Å) | |
| Al | 50000 | 0.617 | 18583 | 0.713 | 1.97% |
| F  | 50000 | 0.333 | 30798 | 0.338 | 6.15% |
| H  | 50000 | 0.249 | 15477 | 0.352 | 3.09% |
| O  | 50000 | 0.538 | 920   | 1.879 | 0.065% |

**Table 2**. DCF algorithm prediction accuracy for the standard and chemical reaction test set of system C. We define $\vec{F}_{MAE} = \langle |\vec{F}_{NN} - \vec{F}_{DFT}| \rangle$. A more complete table of the model's accuracy is shown in the Supplementary Information. Rare event occurrence = (reaction set sample count / full trajectory sample count) for each of the elements. Chemical reaction test set $\vec{F}$ regression does not work well when the rare event occurrence is << 1%.

Going forward, it will be important to increase the fraction of training data set related to atomic events of interest such as chemical reaction. This is very difficult to do in a standard B–P NNFF algorithm because all of the atoms in a simulation frame are coupled by the system total energy. On the other hand, data snapshots in the DCF algorithm are decoupled. This last property is of great interest because in order to improve the prediction accuracy for specific rare atomic event of interest, it is straightforward to simply add atomic environments corresponding to those rare events without having to add significantly more samples which are not physically interesting, a well-known data imbalance problem in the machine learning community.[40] In addition, limiting the force prediction input from just a decoupled local atomic environment may eventually enable researchers to share and mix structure–force training data from different AIMD simulations, eliminating the need for over-redundant force field training data generation. For example, an AIMD training set for ionic diffusion in a bulk electrolyte combined with an AIMD training set for cation dissolution from a cathode surface into the electrolyte may enable large-scale NNFF-MD simulation of cation dissolution from the cathode surface into the bulk electrolyte, a very important phenomenon to study for battery and fuel cell applications.

## Conclusions

We have enabled the direct force vector prediction from local atomic snapshots through the usage of both rotation-invariant and rotation-covariant fingerprint features. This approach enables us to bypass costly fingerprint derivative calculations, typically the computation bottleneck of existing neural network force fields approach, by 180–480×, which may effectively accelerate NNFF-based MD computation speed by similar magnitude. Atomic force projections onto internal covariant axes serve as rotation-invariant hidden states, which were then used to separately train the rotation-invariant and rotation-covariant parts of the network. Finally, transfer learning is used to transfer the NN weights to the full network before further optimization, leading to direct atomic force prediction of a polymer electrolyte, amorphous oxide electrolyte, and oxide surface acid corrosion; practical systems relevant for material discovery and degradation prevention in applications such as batteries and fuel cells.

This force calculator can be directly implemented within a molecular dynamics engine, provided that either a thermostat is used (so total energy need not be conserved) or an efficient neural-network architecture is found that conserves total energy. Beyond atomistic simulation, the rotation-invariant/covariant neural network can be applied to a variety of physical domains to enforce physical symmetries. Future improvements may include more automated fingerprint selection[38] or data representation through the use of generative adversarial networks (GAN)[41] or graph convolutional neural networks,[22] applied separately on the rotation-invariant and covariant components, to further the accuracy of this direct covariant force vector prediction algorithm's accuracy further. Furthermore, true rotation-covariant operation should be implemented for the NN model's component utilizing the covariant features.

## Data availability

The atomic structure–force dataset (for system A, B, and C) and the Python code for training the NNFF of the DCF approach of this work is publicly available at … and can be requested from the corresponding authors.

## Methods

### System A Quantum Data Generation

A low-density MD structure for PEO was first generated and its force field parametrized using Schrodinger Suite and the OPLS 2005 force field. Classical MD was subsequently run in the LAMMPS to compress, anneal, cool, and decompress the structure, bringing it into a PEO state close to equilibrium at operating pressure of $P$ = 1 atm and $T$ = 80 C. The equilibrated structure was then run for 200 ns in an NPT ensemble, and atomic frame snapshots are recorded every 5 ns. Afterward, we extract the local atomic coordinate environment $\{\vec{R}_i\}$ for each of the atoms in the simulation box up to a radius of 10 Å and place them in a large cubic simulation box. DFT simulation was then run using the SIESTA software package,[1] using GGA-PBE functional (and default setting for the rest of the DFT parameters) to calculate the force on central atom $\vec{F}_i$.

### System B & C Quantum Data Generation

System B was a $Li_4P_2O_7$ crystalline system simulated at high temperature via ab-initio molecular dynamics (AIMD). The starting structure was a supercell of $Li_4P_2O_7$ crystal (space group $P2_1/c$) of sixteen formula units, arranged in an orthorhombic simulation box of 10.4 x 14.0 x 16.0 Å with periodic boundary conditions. Simulations were performed using the Vienna Ab-Initio Simulation Package (VASP)[2-4] with a generalized gradient PBE functional,[5] project-augmented-wave (PAW) pseudopotentials,[6] a Γ-point reciprocal-space mesh, and a plane-wave cutoff of 400 eV. The AIMD had a timestep of 2 fs with a Nosé-Hoover thermostat[7] at 3000 K, for a total simulation time of 50 ps.

System C was an alumina surface exposed to HF molecules. The monoclinic cell had β=60°, a=10.5 Å, and c=26.6 Å with periodic boundary conditions. The c-axis contained 10 Å of vacuum with 34 HF molecules; the rest was 32 formula units of $Al_2O_3$ cut along (111). Here too the simulations were performed with VASP, a PBE functional, and PAW pseudopotentials at the Γ point. The plane-wave cutoff was 450 eV. The Nosé thermostat was set at 1000 K and timesteps of 0.5 fs for a total simulation time of 20 ps.

**Cost Benchmarking**

The standard B–P neural networks[8] were trained using the "Atomistic Machine-learning Package" (AMP).[9] Several different architectures were trained using both energies and forces in a joint loss function. The final model used for the evaluation of prediction time was a network with two hidden layers with 50 nodes per layer, using a cutoff radius of 6.5 Å. Once trained, the AMP network was converted into a PROPhet potential file[10] in order to use the efficient, C++-based coupling of PROPhet to the LAMMPS code.[11] The PROPhet standard B–P MD simulation was then run in an NVT simulation at 3000 K for 50 ps at a time step of 1 fs using the Nose-Hoover thermostat.

The DCF NNFF algorithm is similarly trained using cutoff radius of 6.5 Å, two hidden layers with 50 nodes per layer on the rotation-invariant part, and two hidden layers with 40 nodes per layer on the rotation-covariant part. The fingerprint calculation for training and testing are written in Python, while the neural network model calculation is done in Python's TensorFlow module under the Keras framework. To estimate the evaluation time of DCF were it to be implemented within PROPhet's efficient C++ code, we first assume that standard B–P NNFF computational cost in C++ is vastly dominated by the calculation of the fingerprint and its derivatives (a fair assessment given the NN and classical FF overhead shown in **Figure 4**). In DCF, the fingerprint derivative calculation is eliminated. For an average of $m$ = 101 atoms within the 6.5 A cutoff radius for $Li_4P_2O_7$ system, the fingerprint component of DCF is estimated to be 1 / (1 + 3*101) = 0.0033 that of standard NNFF computational cost. Here we have assumed that fingerprint derivative calculation cost is just as expensive as the fingerprint function calculation cost for simplicity, while in reality the fingerprint derivative functional form should be more computationally expensive to calculate. We then added back the LAMMPS and TensorFlow computational overhead from **Figure 4** to obtain the estimate that DCF based MD will be ~127x faster than standard NNFF based MD for system B. This acceleration estimate depends on *m* and should be mostly independent from the size of system B.

**Hidden Markov Model Reaction Snapshot Extraction**

In order to generate the reaction data set for system C, a graph-based approach was used to identify molecules based on atomic connectivity. Interatomic connections are made based on radial cutoff of $R_{ij}^C < 1.4 * (r_i^C + r_j^C)$, where $r_i^C$ and $r_j^C$ are the covalent radii of the chemical species of atom $i$ and $j$.[12] The graph at each time step $G_t$ is represented as a connectivity matrix, which contains connected subgraphs as molecules, located by breadth-first-search. Erroneous connections are made at each snapshot due to thermal vibration and molecular collision. Thus, the presence of each molecule through the trajectory is described by a binary time-series, which is noise-filtered by describing the system as a Hidden Markov Model with observed (connectivity) and hidden states (bonds).[13] The Viterbi algorithm is used to find the filtered signal, which is the most probable hidden state sequence for the Markov chain. The parameters of the HMM are 1) the probability of state change at each time step which was set to a value of 1e-5, 2) the probability of observed state matching the hidden state set to 0.6, and 3) the initial probability of each state which was 0.5. Similar to methods used by Wang and Zheng,[14-15] changes in the connectivity matrix are detected across different frames and changes to subgraphs indicate molecular transformations via chemical reaction. Chemical reactions are verified by balancing reactants and products with the same participating atoms, allowing automated and precise detection of reactivity throughout the course of the simulation.

# Method References

## Acknowledgements

The authors thank S. Falkner and C. Cunha from Bosch Center for Artificial Intelligence (BCAI) for feedback on NNFF algorithm accuracy improvement. This research used resources of the Oak Ridge Leadership Computing Facility at Oak Ridge National Laboratory, which is supported by the Office of Science of the Department of Energy under Contract DE-AC05-00OR22725. The research was partially funded by the Advanced Research Projects Agency – Energy (ARPA-E), U.S. Department of Energy, under Award Number DE-AR0000775.

## Author contributions

J.P. Mailoa conceived and implemented the staggered rotation-invariant and covariant feature separation algorithm. J.P. Mailoa, M. Kornbluth, G. Samsonidze, S.T. Lam, C. Ablitt, and N. Molinari generated the training and test dataset for system A, B, and C. J.P. Mailoa evaluated the NNFF accuracy. J.P. Mailoa, S.L. Batzner, and M. Kornbluth performed the computational cost analysis. S.T. Lam developed the Hidden Markov Model for system C reaction dataset generation. J.P. Mailoa wrote the manuscript. All authors contributed in the manuscript preparation.

## Competing interests

The authors declare no competing interests.

## Materials & Correspondence

Correspondence regarding this manuscript and material requests should be addressed to Jonathan Mailoa at jpmailoa@alum.mit.edu or Boris Kozinsky at bkoz@seas.harvard.edu.

# Electronic Supplementary Information

# Fast Neural Network Approach for Direct Covariant Forces Prediction in Complex Multi-Element Extended Systems

Jonathan P. Mailoa,[1*] Mordechai Kornbluth,[1] Simon L. Batzner,[2,3] Georgy Samsonidze,[1] Stephen T. Lam,[1,2] Chris Ablitt,[1,4] Nicola Molinari,[1,3] and Boris Kozinsky[3,1*]

1) Bosch Research and Technology Center, Cambridge, MA 02139, USA
2) Massachusetts Institute of Technology, Cambridge, MA 02139, USA
3) Harvard School of Engineering and Applied Sciences, Cambridge, MA 02138, USA
4) Imperial College London, London SW7 2AZ, UK

* corresponding author: jpmailoa@alum.mit.edu, bkoz@seas.harvard.edu

# Data Generation using Quantum Mechanics Tools

**System A**

Polyethylene oxide (PEO) system consisting of 10 polymer chains 100 monomer each was built and equilibrated using classical OPLS 2005 force field designed for organic chemistries in LAMMPS software package. There are 7090 atoms in the system (2020 C atoms, 4060 H atoms, and 1010 O atoms). Classical MD frame snapshots were outputted every 5 ns of NPT ensemble run, and 40 frames were outputted. Out of those 40 frame snapshots, spherical atomic snapshots with cutoff radius of 10 Å were extracted (40400 C, 40600 H, and 44440 O atomic snapshots were chosen). After terminating the open bonds with hydrogen, SIESTA DFT software package was used on these 10 Å radius structures to generate the force on central atom. The training and test dataset are then taken from these atomic snapshots.

**System B**

$Li_4P_2O_7$ structure with 208 atoms (64 Li, 112 O, and 32 P atoms) was run using ab-initio MD in VASP software package. A frame snapshot was outputted every 2.0 fs during a 50 ps ab-initio MD run, leading to the generation of 25000 frame snapshots (5.2 million atomic snapshots). Out of those, 750000 atomic snapshots were randomly chosen for each of Li, O, and P atoms. The training and test dataset are then taken from these atomic snapshots.

**System C**

$Al_2O_3$ surface structure in the presence of HF molecules with 228 atoms (64 Al, 34 F, 34 H, and 96 O atoms) was run using ab-initio MD in VASP software package. A frame snapshot was outputted every 0.5 fs during a 7.3685 ps ab-initio MD run, leading to the generation of 14737 frame snapshots (3360036 atomic snapshots). Out of those, 500000 atomic snapshots were randomly chosen for each of Al, F, H, and O atoms. The training and test dataset are then taken from these atomic snapshots.

# Hyperparameter for $\bar{G}$ and $\{\vec{A}\}$ Fingerprints

First we outline the fingerprint hyperparameters we used to generate direct NNFF models for system A, B and C in the main text's **Table 1**. The relevant 2-body and 3-body functions for $\bar{G}$ and $\{\vec{A}\}$ and are listed in equation (2-8):

$$f_c(r) = \begin{cases} 0.5 \cdot [\cos(\pi r/R_c) + 1] & \text{for } r \leq R_c \\ 0 & \text{for } r > R_c \end{cases} \quad (1)$$

$$G_1^{el1}(\{\vec{R}_i\}) = \sum_{j \in el1} f_c(|\vec{r}_{ij}|) \quad (2)$$

$$G_2^{el1}(\{\vec{R}_i\}) = \sum_{j \in el1} e^{-\eta(|\vec{r}_{ij}|-R_s)^2} f_c(|\vec{r}_{ij}|) \quad (3)$$

$$G_3^{el1}(\{\vec{R}_i\}) = \sum_{j \in el1} \cos(\kappa |\vec{r}_{ij}|) f_c(|\vec{r}_{ij}|) \quad (4)$$

$$G_4^{el1,el2}(\{\vec{R}_i\}) = 2^{1-\zeta} \sum_{\substack{j \in el1 \\ k \in el2 \\ j \neq k}} (1 + \lambda \cos\theta_{ijk})^\zeta e^{-\eta(|\vec{r}_{ij}|^2 + |\vec{r}_{ik}|^2 + |\vec{r}_{jk}|^2)} f_c(|\vec{r}_{ij}|) f_c(|\vec{r}_{ik}|) f_c(|\vec{r}_{jk}|) \quad (5)$$

$$G_5^{el1,el2}(\{\vec{R}_i\}) = 2^{1-\zeta} \sum_{\substack{j \in el1 \\ k \in el2 \\ j \neq k}} (1 + \lambda \cos\theta_{ijk})^\zeta e^{-\eta(|\vec{r}_{ij}|^2 + |\vec{r}_{ik}|^2)} f_c(|\vec{r}_{ij}|) f_c(|\vec{r}_{ik}|) \quad (6)$$

$$\vec{V}_1^{el1}(\{\vec{R}_i\}) = \sum_{j \in el1} \vec{r}_{ij} e^{-\eta(|\vec{r}_{ij}|-R_s)^2} f_c(|\vec{r}_{ij}|) \quad (7)$$

$$\vec{A}_1^{el1}(\{\vec{R}_i\}) = \begin{cases} 0 & \text{for } \vec{V}_1^{el1} = 0 \\ \vec{V}_1^{el1}/|\vec{V}_1^{el1}| & \text{otherwise} \end{cases} \quad (8)$$

| Center Atom | Function | Parameters | Count |
|---|---|---|---|
| **Rotation-Invariant Features** | | | |
| C, H, O | $G_1^C, G_1^H, G_1^O$ | $R_c = 6$ | 3 |
| | $G_2^C, G_2^H, G_2^O$ | $R_c = 6, \eta = [0.5, 1.0, 1.5]$, $R_s = [1.0, 2.0, 3.0, 4.0, 5.0]$ | 45 |
| | $G_2^C, G_2^H, G_2^O$ | $R_c = 6, \eta = 1000, R_s = [1.0, 1.1, 1.2, 1.3, 1.4, 1.5]$ | 18 |
| | $G_2^C, G_2^H, G_2^O$ | $R_c = 6, \eta = 100, R_s = [2.0, 2.2, 2.4, 2.6]$ | 12 |
| | $G_3^C, G_3^H, G_3^O$ | $R_c = 6, \kappa = [1.0, 2.0, 3.0, 4.0, 5.0]$ | 15 |
| | $G_4^{C,C}, G_4^{C,H}, G_4^{C,O}, G_4^{H,H}, G_4^{H,O}, G_4^{O,O}$ | $R_c = 6, \zeta = [1, 2, 4, 16, 64], \lambda = [1, -1]$, $\eta = [0.01, 0.05, 0.1, 0.2, 0.5, 1.0, 2.0, 5.0]$ | 480 |
| | $G_5^{C,C}, G_5^{C,H}, G_5^{C,O}, G_5^{H,H}, G_5^{H,O}, G_5^{O,O}$ | $R_c = 6, \zeta = [1, 2, 4, 16, 64], \lambda = [1, -1]$, $\eta = [0.01, 0.05, 0.1, 0.2, 0.5, 1.0, 2.0, 5.0]$ | 480 |
| **Rotation-Covariant Features** | | | |
| C | $\vec{A}_1^C$ | $R_c = 6, \eta = 100, R_s = [1.5, 2.4]$ | 6 |
| | $\vec{A}_1^H$ | $R_c = 6, \eta = 100, R_s = [1.0, 1.1, 1.2]$ | 9 |
| | $\vec{A}_1^O$ | $R_c = 6, \eta = 100, R_s = [1.4, 2.4]$ | 6 |
| H | $\vec{A}_1^C$ | $R_c = 6, \eta = 100, R_s = [1.1, 2.1]$ | 6 |
| | $\vec{A}_1^H$ | $R_c = 6, \eta = 100, R_s = [1.7, 2.2]$ | 6 |
| | $\vec{A}_1^O$ | $R_c = 6, \eta = 100, R_s = [1.9, 2.1]$ | 6 |
| O | $\vec{A}_1^C$ | $R_c = 6, \eta = 100, R_s = [1.3, 1.5. 2.05, 2.2]$ | 12 |
| | $\vec{A}_1^H$ | $R_c = 6, \eta = 100, R_s = [1.95, 2.0, 2.1, 2.2]$ | 12 |
| | $\vec{A}_1^O$ | $R_c = 6, \eta = 100, R_s = [2.0, 2.2]$ | 6 |

**Table S1.** Selected rotation-invariant and rotation-covariant fingerprint features for system A (**Table 1** in the main text), amorphous polyethylene oxide polymer (PEO). NN1 consists of 2 hidden layers, 120 nodes each for each $F_{int,i}$ node. NN2 consists of 2 hidden layers, 40 nodes each for each $F_i$ node.

| Center Atom | Function | Parameters | Count |
|---|---|---|---|
| **Rotation-Invariant Features** | | | |
| Li, O, P | $G_2^{Li}, G_2^O, G_2^P$ | $R_c = 6.5, \eta = [0.05, 4, 20, 80],$ $R_s = [0, 1, 3]$ | 36 |
| | $G_4^{Li,Li}, G_4^{Li,O}, G_4^{Li,P},$ $G_4^{O,O}, G_4^{O,P}, G_4^{P,P}$ | $R_c = 6.5, \zeta = [1, 4, 16], \lambda = [1, -1],$ $\eta = [0.005, 0.02, 0.1]$ | 108 |
| O | $G_2^{Li}, G_2^O, G_2^P$ | $R_c = 6.5, \eta = 1000,$ $R_s = [1.4, 1.6, 1.8, 2.0, 2.5, 2.7, 3.0, 3.3]$ | 24 |
| **Rotation-Covariant Features** | | | |
| Li, O, P | $\vec{A}_1^{Li}, \vec{A}_1^O, \vec{A}_1^P$ | $R_c = 6.5, \eta = 1.0, R_s = 0$ | 9 |
| | $\vec{A}_1^{Li}, \vec{A}_1^O, \vec{A}_1^P$ | $R_c = 6.5, \eta = 2.0, R_s = 1.0$ | 9 |
| O | $\vec{A}_1^{Li}$ | $R_c = 6.5, \eta = 10, R_s = 1.4$ | 3 |
| | $\vec{A}_1^O$ | $R_c = 6.5, \eta = 10, R_s = 2.0$ | 3 |
| | $\vec{A}_1^P$ | $R_c = 6.5, \eta = 100, R_s = [1.6, 1.8]$ | 6 |

**Table S2**. Selected rotation-invariant and rotation-covariant fingerprint features for system B (**Table 1** in the main text), amorphous $Li_4P_2O_7$. NN1 consists of 2 hidden layers, 50 nodes each for each $F_{int,i}$ node. NN2 consists of 2 hidden layers, 40 nodes each for each $F_i$ node.

| Center Atom | Function | Parameters | Count |
|---|---|---|---|
| **Rotation-Invariant Features** | | | |
| Al, F, H, O | $G_2^{Al}, G_2^F, G_2^H, G_2^O$ | $R_c = 6.5, \eta = [0.05, 4, 20, 80],$ $R_s = [0, 1, 3]$ | 48 |
| | $G_2^{Al}, G_2^F, G_2^H, G_2^O$ | $R_c = 6.5, \eta = 1000,$ $R_s = [1.0, 1.7, 1.8, 1.9, 2.0]$ | 20 |
| | $G_4^{Al,Al}, G_4^{Al,F}, G_4^{Al,H}, G_4^{Al,O},$ $G_4^{F,F}, G_4^{F,H}, G_4^{F,O}, G_4^{H,H},$ $G_4^{H,O}, G_4^{O,O}$ | $R_c = 6.5, \zeta = [1, 4, 16], \lambda = [1, -1],$ $\eta = [0.005, 0.02, 0.1]$ | 180 |
| Al | $G_2^{Al}, G_2^F, G_2^H, G_2^O$ | $R_c = 6.5, \eta = 200,$ $R_s = [2.2, 2.3, 2.4, 2.5]$ | 16 |
| **Rotation-Covariant Features** | | | |
| Al | $\vec{A}_1^{Al}$ | $R_c = 6.5, \eta = 5.0, R_s = [2.3, 2.7, 3.1]$ | 9 |
| | $\vec{A}_1^F$ | $R_c = 6.5, \eta = 50, R_s = [1.8, 1.9]$ | 6 |
| | $\vec{A}_1^H$ | $R_c = 6.5, \eta = 50, R_s = [2.4, 2.6, 2.8]$ | 9 |
| | $\vec{A}_1^O$ | $R_c = 6.5, \eta = 50, R_s = [1.8, 1.95, 2.1]$ | 9 |
| F, H | $\vec{A}_1^{Al}, \vec{A}_1^F, \vec{A}_1^H, \vec{A}_1^O$ | $R_c = 6.5, \eta = 100,$ $R_s = [1.0, 1.6, 1.8, 2.0]$ | 48 |
| O | $\vec{A}_1^{Al}$ | $R_c = 6.5, \eta = 100, R_s = [1.7, 1.9, 2.1]$ | 9 |
| | $\vec{A}_1^F$ | $R_c = 6.5, \eta = 50, R_s = [2.3, 2.7, 3.1]$ | 9 |
| | $\vec{A}_1^H$ | $R_c = 6.5, \eta = 50, R_s = 1.0$ | 3 |
| | $\vec{A}_1^O$ | $R_c = 6.5, \eta = 50, R_s = [1.5, 2.4, 2.8, 3.2]$ | 12 |

**Table S3**. Selected rotation-invariant and rotation-covariant fingerprint features for system C (**Table 1** in the main text), $Al_2O_3$ surface – HF reaction. NN1 consists of 2 hidden layers, 50 nodes each for each $F_{int,i}$ node. NN2 consists of 2 hidden layers, 40 nodes each for each $F_i$ node.

| Center Atom | Function | Parameters | Count |
|---|---|---|---|
| **Rotation-Invariant Features** | | | |
| Li, O, P | $G_2^{Li}, G_2^{O}, G_2^{P}$ | $R_c = 6.5, \eta = [0.05, 4, 20, 80], R_s = 0$ | 12 |
| | $G_4^{Li,Li}, G_4^{Li,O}, G_4^{Li,P}, G_4^{O,O}, G_4^{O,P}, G_4^{P,P}$ | $R_c = 6.5, \zeta = [1,4], \lambda = [1,-1], \eta = 0.005$ | 24 |
| **Rotation-Covariant Features** | | | |
| Li, O, P | $\vec{A}_1^{Li}, \vec{A}_1^{O}, \vec{A}_1^{P}$ | $R_c = 6.5, \eta = 1.0, R_s = 0$ | 9 |
| | $\vec{A}_1^{Li}, \vec{A}_1^{O}, \vec{A}_1^{P}$ | $R_c = 6.5, \eta = 2.0, R_s = 1.0$ | 9 |

**Table S4**. Selected rotation-invariant and rotation-covariant fingerprint features for system B (**Figure 4** in the main text), amorphous Li$_4$P$_2$O$_7$. Rotation-invariant features are used by both standard B–P and DCF NNFF algorithm, while the rotation-covariant features are only used by the DCF NNFF algorithm. In the standard B – P algorithm, 2 hidden layers with 50 nodes on each layer are used. In the DCF algorithm, NN1 consists of 2 hidden layers, 50 nodes each for each $F_{int,i}$ node, while NN2 consists of 2 hidden layers, 40 nodes each for each $F_i$ node.

# Quantum Force Information Encoded within $\bar{G}$ and $\{\vec{A}\}$

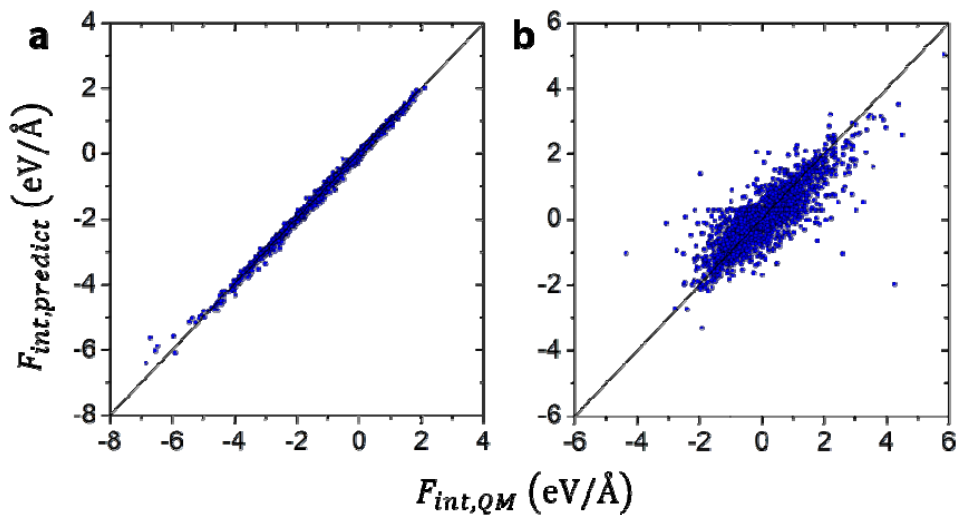

**Figure S1**. $\bar{F}_{int}$ component prediction by NN1 for H atoms in system A. Examples are shown for: a) well-chosen $\vec{A}$ and b) less relevant $\vec{A}$.

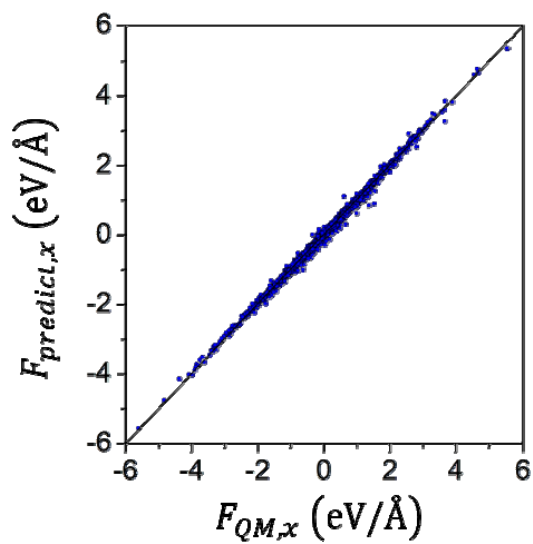

**Figure S2**. $\vec{F}$ component prediction by NN2 for H atoms in system A using a relevant set of $\{\vec{A}\}$.

## DCF Force Prediction Evaluation on Standard Test Set

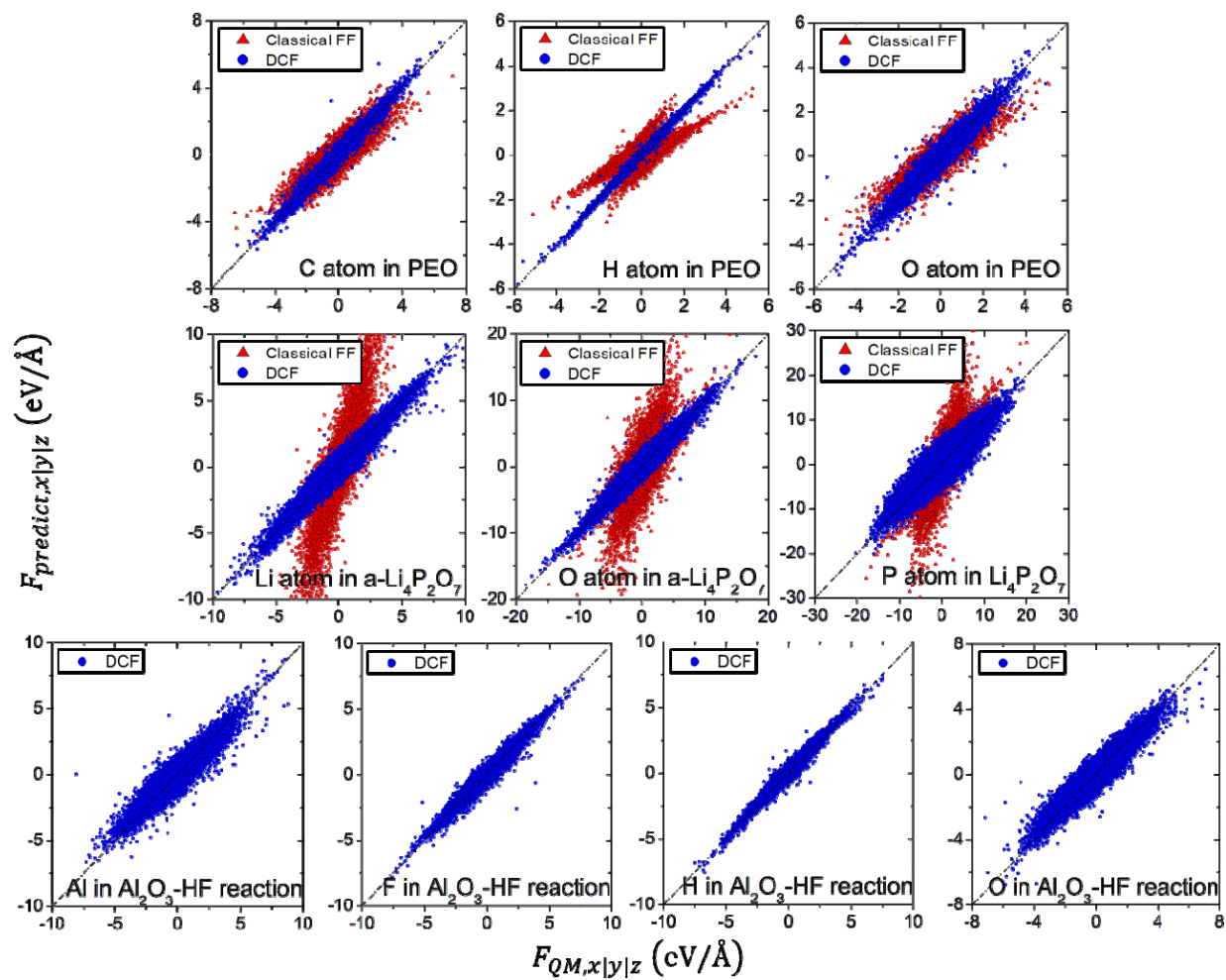

**Figure S3**. DCF force prediction correlation plot on the standard test set for system A, B, and C. Classical FF prediction is shown when available.

| System | Atom | Classical FF $\vec{F}_{MAE}$ (eV/Å) | Classical FF $\vec{F}_{RMSE}$ (eV/Å) | NNFF $\vec{F}_{MAE}$ (eV/Å) | NNFF $\vec{F}_{RMSE}$ (eV/Å) | DFT $\langle|\vec{F}_{DFT}|\rangle$ (eV/Å) | Classical FF $\frac{\vec{F}_{MAE}}{\langle|\vec{F}_{DFT}|\rangle}$ | Classical FF $\frac{\vec{F}_{RMSE}}{\langle|\vec{F}_{DFT}|\rangle}$ | NNFF $\frac{\vec{F}_{MAE}}{\langle|\vec{F}_{DFT}|\rangle}$ | NNFF $\frac{\vec{F}_{RMSE}}{\langle|\vec{F}_{DFT}|\rangle}$ | Training Count | Test Count |
|---|---|---|---|---|---|---|---|---|---|---|---|---|
| PEO | C | 1.222 | 1.278 | 0.361 | 0.422 | 2.365 | 52% | 54% | 15% | 18% | 36360 | 4040 |
| | H | 1.088 | 1.126 | 0.128 | 0.149 | 1.411 | 77% | 80% | 9% | 11% | 36540 | 4060 |
| | O | 0.93 | 1.009 | 0.431 | 0.499 | 1.736 | 54% | 58% | 25% | 29% | 39996 | 4444 |
| Li$_4$P$_2$O$_7$ | Li | 2.736 | 4.43 | 0.603 | 0.662 | 1.629 | 168% | 272% | 37% | 41% | 675000 | 75000 |
| | O | 5.521 | 6.296 | 0.864 | 0.973 | 3.446 | 160% | 183% | 25% | 28% | 675000 | 75000 |
| | P | 4.919 | 5.718 | 2.375 | 2.643 | 5.868 | 84% | 97% | 40% | 45% | 675000 | 75000 |
| Al$_2$O$_3$- HF | Al | N/A | N/A | 0.617 | 0.704 | 1.846 | N/A | N/A | 33% | 38% | 450000 | 50000 |
| | F | | | 0.333 | 0.381 | 1.525 | | | 22% | 25% | 450000 | 50000 |
| | H | | | 0.249 | 0.289 | 1.116 | | | 22% | 26% | 450000 | 50000 |
| | O | | | 0.544 | 0.611 | 1.551 | | | 35% | 39% | 450000 | 50000 |

**Table S4.** Complete evaluation of force prediction accuracy for the new DCF algorithm. We define $\vec{F}_{MAE} = \langle|\vec{F}_{NN} - \vec{F}_{DFT}|\rangle$ and

$$\vec{F}_{RMSE} = \sqrt{\langle|\vec{F}_{NN} - \vec{F}_{DFT}|^2\rangle}.$$

# DCF Force Prediction Evaluation on System C Reaction Test Set

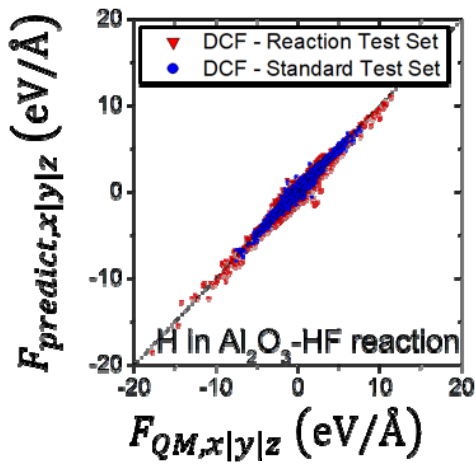

**Figure S4**. DCF force prediction correlation plot for H atoms in the reaction test set of system C (red triangle), showing slightly worse performance than the prediction on the standard test set (blue dots).

| System | Atom | $\vec{F}_{MAE}$ (eV/Å) | $\vec{F}_{RMSE}$ (eV/Å) | $\langle|\vec{F}_{DFT}|\rangle$ (eV/Å) | $\dfrac{\vec{F}_{MAE}}{\langle|\vec{F}_{DFT}|\rangle}$ | $\dfrac{\vec{F}_{RMSE}}{\langle|\vec{F}_{DFT}|\rangle}$ |
|---|---|---|---|---|---|---|
| Al$_2$O$_3$–HF reaction test set | Al | 0.713 | 0.841 | 2.216 | 32% | 38% |
| | F | 0.338 | 0.412 | 1.71 | 20% | 24% |
| | H | 0.352 | 0.424 | 1.64 | 21% | 26% |
| | O | 1.879 | 4.231 | 1.791 | 105% | 236% |

**Table S5**. DCF algorithm prediction accuracy for the standard and chemical reaction test set of system C.

We define $\vec{F}_{MAE} = \langle|\vec{F}_{NN} - \vec{F}_{DFT}|\rangle$ and $\vec{F}_{RMSE} = \sqrt{\langle|\vec{F}_{NN} - \vec{F}_{DFT}|^2\rangle}$.

# Large Force Error Propagation in Pseudoinverse Method

During the step of performing $(\bar{F}_{int,i}, \{\vec{A}_i\}) \rightarrow \vec{F}_i$ prediction, one analytical method which has been proposed is the usage of pseudoinverse.[1] In this pseudoinverse method, one takes pseudoinverse of the relationship $\bar{F}_{int,i} = \{\vec{A}_i\} \cdot \vec{F}_i$ to calculate $\vec{F}_i = \left(\{\vec{A}_i\}^T \{\vec{A}_i\}\right)^{-1} \{\vec{A}_i\}^T \cdot \bar{F}_{int,i}$. In ideal situations when $\{\vec{A}_i\}$ spans the full orthogonal 3-dimensional space and there is small error on $\bar{F}_{int,i}$ prediction, this pseudoinverse method can analytically determine $\vec{F}_i$. However, in practice there is no guarantee that $\{\vec{A}_i\}$ will span the entire 3-dimensional space. In those instances, the matrix inversion tends to diverge and the $\vec{F}_i$ pseudoinverse function tends to significantly amplify errors in the $\bar{F}_{int,i}$ prediction.